# Creating Publishing Accounts for University Professors on Global Scientific Websites (ORCID, Research Gate, Google Scholar)

Prof. Dr. Ahmed Shaker Alalaq

University of Kufa - College of Arts

**Research Summary**

For Iraqi university professors, the issue of electronic publishing and creating electronic accounts on reputable international research sites has always been an important matter related to the educational aspect and its sobriety, and the extent to which the strength of universities is measured by the strength of research published on global platforms and repositories, and this is a quality indicator

In our research, we will discuss the most prominent international sites and repositories that receive the research and articles of researchers and teachers, which are considered a basic pillar of building the educational process in Iraqi universities. Although these sites are overlooked by many, they are considered a necessity for any university professor and an interface that we can call (a media interface). ) To promote the teaching research papers and thus give an impression to the recipient and the evaluator of the extent of the university's scientific strength and the ability of its professors at the level of scientific publishing

Perhaps among the most prominent sites on which we always encourage professors to create accounts are (ORCID), (Reserach Gate), and (Google Scholar), and how to publish and promote their research through social media or through educational platforms, conferences, and scientific workshops

Then we try to explain, in the course of the research, in a smooth manner, the ways to activate accounts on these platforms, supported by pictures and a comprehensive step-by-step explanation, as a gesture to encourage the spread of the culture of electronic publishing in light of the escalation of the digital and computing revolution and the desire to catch up with its accelerating pace

Keywords (documentation, accounts, programs, quotation, creation, SQL, Orchid, Research Kit)

Introduction

In light of the accelerating pace of the digital revolution the world is witnessing today, and the increasing pace of institutional university work to obtain advanced positions in global university rankings and technical and technological indicators, and in light of the growing desire of many university professors to obtain the greatest possible access to their research and research papers and to facilitate communication with each other, it has become necessary for everyone concerned to create multiple accounts on global digital repositories and websites where they can publish their articles and research. Naturally, there are many such sites that facilitate this process.

In our research, we will shed light on these sites with some simplicity and ease so that everyone can create these accounts and then publish research and scientific paper abstracts.

Research Objectives:

The research aims to achieve a set of important matters related to the work of the university professor, including:

1-Facilitating and simplifying the methods of creating official accounts for them on scientific repositories and websites.

2-Publishing their research papers and articles so that they can be accessed by the largest possible number of researchers, interested parties, and specialists.

3-The possibility of promoting these researches and summaries to help increase scientific citations.

4-Reviewing data and studies related to the researcher's specialization in various world languages, and then benefiting from what the specialists have written.

5-Diversifying the use of sources and references and keeping up with the latest developments in scientific research.

**Axis One: The Importance of ORCID Account and its Features for University Professors**

The ORCID website is one of the most prominent scientific electronic websites for university professors. It is an abbreviation for (Open Researcher and Contributor ID). It consists of symbols and numbers to distinguish between scientists and academic authors. That is, each researcher has a special identification number (ID) that distinguishes him from other researchers. It is a non-profit global organization that collects fees from our member organizations. These members are a wide range of research centers and official global universities and some publishing houses. ORCID provides a range of free services to researchers, including:
• A unique and persistent identifier for researchers (ID) ORCID
• A permanent record of research and scientific papers added by researchers to their personal pages.
• A set of applications and an application interface for a set of support data and technical support.

The website's interface is very easy to use and everyone can interact with the data provided on it. The website supports many languages, including Arabic.

After logging in via the link below, which is the official website: https://orcid.org/

You are asked to (register) (new registration) and fill in your data, which are: your full name in either Arabic or English, and your official email. Then you move to the second step, which includes a set of data related to notifications and the protection of your account, the most important of which is (entering a password). It is preferable that the password be a combination of symbols, numbers, and letters. After that, you move to the third stage, the offer, and conditions, including choosing who can see your personal file, and it is best to make it a visible file for everyone and then click on agreeing to all the terms of the platform. These are all very easy procedures that do not require any effort.

After completing the registration process, the main interface of your account appears, which contains a set of editable data, which are:

• Your personal name that you want to appear to the public.

• Personal biography and your scientific profile: You can add your personal and scientific biography in both Arabic and English and make it visible to the public.

• Employment: This refers to your place of work or the places where you have worked to inform researchers about your research interests and places of work.

• Education and qualifications: Add certificates and the level of education obtained.

• Professional activities: You can add your invited positions or memberships, awards or rewards you have received, and contributions made to the service of organizations                            or                            institutions

• Works: This is the most prominent and important option on the (ORCID) website, and you can add all your research and research papers. The beauty of this option is that it shows you the author's name, the date the research was added, the last modification to it, and how to cite it (Citation). You can also add the research or research paper by adding its (DOI), in which case the visitor to your account can, by clicking on the (DOI) ID, go to the original page of the research site, which I mean the website of the journal that published your research.

• You can add links to your accounts on social networking sites, which makes it easier for researchers to reach you and communicate with you, and thus facilitate the exchange of knowledge and science between researchers from different                countries                of                the                world. Through (ORCID) you can get an ID that distinguishes you, as we said, from all researchers, and at the same time it is a site for social, cognitive, and scientific communication. It introduces specialists to your scientific and cognitive products

## Axis Two: The Importance of Research Gate for University Professors

ResearchGate is one of the electronic websites that provides many scientific services to researchers from various scientific disciplines. It is like a database for research and articles with open access for free. These researches were put by researchers specializing in their fields of work, and it is possible to view this vast amount of scientific studies from different languages, download them, and use them as sources in writing scientific research while referring to their original authors. The site has been operating since 2008, and according to the official website, the number of researchers registered on it has exceeded 14 million     from     all     countries     of     the     world.     (uobaghdad). The site provides a set of free services that enhance the value of sound scientific research, and perhaps among them is the creation of an official account for everyone for free by using the researcher's institutional email (electronic mail) as well as personal email, and naming the account with the researcher's full name and workplace, and thus the possibility of opening the door to communication, dialogue, and exchange of information and science between researchers from different countries of the world. It is also possible, through the ResearchGate account, to ask scientific questions and publish them on the researcher's official page, and other researchers can answer the question, which matures the idea of obtaining information, exchanging ideas, and presenting opinions with different orientations, ensuring free access to that information away from restrictions. (Alalaq, 2023). What distinguishes the Research Gate website is that the researcher can upload his research in the form of multiple types of files (pdf, word), even in the form of an image, poster, or charts, and other files. It is also possible to do a (search) for your research papers published on the pages of scientific journals - if the journal indexes its research in Research Gate - or do a search for your name, then all your research will appear, and from the add-on service, the research will be added to your account on the site. It is possible to make these files available to everyone in general and benefit from them, view them, comment on them, and express opinions and observations. The site provides another service, which is communicating with researchers through messages and sending files of various kinds, and thus the researcher can obtain whatever he wants from research and files by communicating with the original author of the research. (Wikipedia).

The site also provides a "citation" service for all your research and to know the research that has used your research and quoted from it, which in itself constitutes one of the distinguished scientific services that the researcher and the institution in which he works need, and to know the global classifications of his research and his institution alike, The more the number of citations for your research increases, the more the research becomes solid and widespread. It is also possible to know the number of readings that your publications have obtained. The more the readings increase, the greater the researcher's audience and the more researchers know about him and his scientific research. Here, it is necessary to refer to a very important point, which is that the researcher's scientific biography must be attached in both Arabic and English languages so that researchers can view his scientific and intellectual products, orientations, and research path. It is also necessary to attach all his accounts on social media, especially Google Scholar and Scopus accounts for the same purpose. (Al-Sabit, D.                                                   -                                                   T).

The site provides accurate statistics and weekly charts of the number of researches cited in research papers and the number of published researches, and thus the more the citations increase, the higher the (H-index) percentage, similar to the Google Scholar account or the Scopus account. The site also provides statistics on the degree of research and scientific operations taking place on his account according to the name (RI Score) (research interests points), so the more his research and activity on the platform increase, the higher his (RI) degree                                    will                                    be.

The site provides the possibility of creating (followers) for researchers and making common research interests, and the more the number of followers increases, the more opportunities for publishing your research worldwide increase, and thus increase the citations and scientific references to your research. The site also provides a service for saving research for other researchers that the researcher may need to read and review in the future. From here, I invite all researchers to create accounts on ResearchGate and upload all their research, research papers, and scientific documents. (Alalaq, 2023).

Once you enter one of the browsers, we write (Research Gate) or go to the main link of the site (https://www.researchgate.net/), the phrase (registration is free) appears.

**Ways to Increase Research Gate Profile Visibility**

Research Gate offers a range of free features and services that researchers can utilize to boost their profile visibility, increase the readership (reads) and citations of their publications, and enhance their Research Interest Score (RIS), a measure of research engagement on ResearchGate. Some key strategies include:

• Complete your profile: Include a clear and professional photo, accurate contact information, and a concise bio in both English and your native language. This makes it easier for others to connect with you.

• Ask and answer questions: Engage in the Questions section by posing research queries or initiating discussions. This can lead to valuable insights, feedback, and potential collaborations.

• Explore job opportunities: Utilize the Jobs section to discover relevant job openings and connect with potential employers. This provides an opportunity to showcase your research and establish professional connections.

• Add publications: Upload your research papers, presentations, and other materials. Include the Digital Object Identifier (DOI) for easy access and to protect your copyright.

• Increase followers: The more followers you have, the more exposure your research will receive. Follow other researchers in your field and engage with them through messaging.

• Highlight research: Use the Research Spotlight feature to draw attention to specific publications or research projects. This allows you to present important findings or promote ongoing work.

• Confirm authorship: Search for and claim your published papers to ensure proper attribution. This helps consolidate your research output and establish your                                                                      credibility.
• Follow relevant journals: Follow journals that publish in your field of interest. This simplifies adding new publications to your profile and keeps you updated on            the            latest            research.
• Provide full-text access: Make your publications openly available by uploading PDFs or Word documents. This increases their accessibility and visibility.
• Share your profile: Promote your ResearchGate profile on social media and other platforms. This helps disseminate your research and connect with a wider audience.

Understanding Research Interest Score (RIS)

The RIS is a metric that helps researchers track the impact of their research on ResearchGate. It combines factors such as the number of views, citations, and recommendations received for publications.

RIS can take time to accumulate, depending on the researcher's activity on the platform. Adding new publications, engaging in discussions, and receiving citations all contribute to increasing RIS. ResearchGate emphasizes transparency in the calculation of RIS, ensuring that it remains a reliable indicator of research engagement.

The RIS calculation formula is as follows:

Partial views (viewing only the abstract): 0.05

Full views: 0.15

Recommendations: 0.25

Citations: 0.5

To enhance RIS, researchers should continue publishing and updating their work on ResearchGate and promote their research through social media to increase readership.

## Chapter Three: Google Scholar Website, Its Importance, and How to Create an Account

Google Scholar is one of the most prominent scientific repositories and engines that provides researchers and specialists with millions of open-source scientific materials. It is owned by Google. It can be fully utilized by creating a personal account for each researcher. Through the account, it is possible to attach all research and scientific papers specific to researchers on their official accounts. Accordingly, many can view these research papers and benefit from the information and scientific content they contain.

As for the search methods, they are diverse in this scientific engine. It is possible to search by author's name or by the title of the article or book in either Arabic or English. It is also possible to search by the name of the publishing house, the name of the scientific journal, or its standard number.

Upon entering the official website of the engine by clicking on the link below:

https://scholar.google.com/

It asks you to log in if you wish to create a personal account, and if you do not, there is no need to. The engine provides everyone with the feature of searching for sources and research papers in various specializations and languages and with full texts without requesting registration in the repository. This in itself is one of the most prominent and important features that distinguish this repository from other search engines, some of which require creating an account. (sciegate).

Those wishing to create an account, who are researchers and academics, are required to register via the institutional mail so that it can be verified by sending a verification message to the registered email. After that, the features provided by the engine can be summarized (Abdul Salam, Dr. -T).

Create and organize a personal file: Through this option, you can create a personal file and modify it and add the name you want in the language you want. The personal file also includes adding your personal website and some description that you want others to see.

My Library option: This is an option that allows researchers to include the articles and books that they wish to view and benefit from later, once we add those articles to this option.

Metrics: This is an option that allows researchers to view the most important international publishing entities that have been cited during the past five years and are organized according to the most cited.

Notifications option: This is an option through which researchers can create notifications specific to their articles or articles that they wish to view if they are available on the site or notifications specific to their personal file and modify it or add articles to their file and account or a notification regarding citations from his articles and other alerts controlled by the researcher.

Settings option: This is the most important option available on the engine. Through it, you can control the display of results and the number of drop-down pages in the search. You can also specify the researcher's language or the language of the articles. You can also delete the account completely through the settings option. This is something I do not recommend.

The repository also provides a quotation service "citation" for research and studies by entering the title of the research or study in the (Research) box as shown in the picture:

Then choose the research for which you want to quote and click on the "Quote" option located below the research title. Then, different forms of quotation appear, and then we choose the form and type of quotation we want, such as (APA) or (MLA) or (ISO). After copying the text, we go and paste it into the list of sources and references for our new research, which we want to publish. (Al-Qaed, 2014).

Conclusion

The research focused on a set of data and information in which we discussed the methods of creating official accounts on global scientific websites and repositories that all researchers and specialists need. The research aimed to shed light on a number of websites that include hundreds of thousands of research papers and scientific abstracts in various languages, which help researchers and specialists to keep abreast of the latest developments and aspirations of scientific events that occur here and there.

The issue of creating scientific accounts and publishing research for university professors is one of the necessities of any professor's work. It is an indicator of his and the projects of his studies and scientific research. It is also a way of conveying information to the recipient in a modern practical way and makes it easy for everyone to view and read those papers and know what is going on in them of ideas in the simplest and easiest ways.

Today, universities measure the scientific indicators of their professors and their attainment of advanced scientific centers by the extent to which these accounts are created, the number of research published in them, and the number of citations obtained by those research. Thus, the number of visits to those accounts is calculated. The more these visits increase, the more it gives an indication of the strength of those universities and the extent of the researchers' need for what those accounts contain in terms of solid scientific research.

# References

Alalaq, Ahmed. (2023). Its features for the university professor (ORCID) are important calculations. 10.13140/RG.2.2.15433.31844.

https://2u.pw/bOEeQ16

https://copew.uobaghdad.edu.iq/?page_id=31864

Tahani Muhammad Al-Sabeet, The Researcher's Guide to Using the Search Engine Research Kit, D-T), Umm Al-Qura University, https://www.slideshare.net/DSR_UQU/researchgate-guide

Alalaq, Ahmed. (2023). (Reserach Gate)Methods to increase account activation. 10.13140/RG.2.2.22897.86888.

Alalaq, Ahmed. (2023). The university professor (Reserach Gate) is important. 10.13140/RG.2.2.36292.81285/1.

Alalaq, A. (2016). History of the development of journalism in Iran 1819-1914 AD. *Kufa Journal of Arts*, *1*(25), 149–190. https://doi.org/10.36317/kaj/2015/v1.i25.6301

Muhammad, A., & Alalaq, A. (2012). The Iranian Kurdistan Democratic Party 1963-1979 AD. *Kufa Journal of Arts*, *1*(14), 100–124. https://doi.org/10.36317/kaj/2012/v1.i14.6282

Alalaq, A. (2021). The impact of Morgan Schuster's financial mission in Iran 1910-1912 AD, a historical study. *Kufa Journal of Arts*, *1*(39), 305–340. https://doi.org/10.36317/kaj/2019/v1.i39.788

Alalaq, A. (2021). History of the development of journalism in Iran 1819-1914 AD. *Kufa Journal of Arts*, *1*(38), 315–360. https://doi.org/10.36317/kaj/2018/v1.i38.758

Alalaq, A. (2021). Iran's Freedom Renaissance Party and its Impact on the Contemporary History of Iran 1979-1981 A.D. An Analytical Study of the Party's Statements. *Kufa Journal of Arts*, *1*(46), 163–210.
https://doi.org/10.36317/kaj/2021/v1.i46.647

Alalaq, A. (2015). The events of Black September in 1970 AD in the light of the Kissinger-Nixon correspondence. *Kufa Journal of Arts*, *1*(24), 257–296.
https://doi.org/10.36317/kaj/2015/v1.i24.6308
al-Hakim, A., & Alalaq, A. (2016). The position of the Iranian press on the internal developments in Iran 1918-1925 AD in the light of Persian documents, a study in models. *Kufa Journal of Arts*, *1*(26), 267–302.
https://doi.org/10.36317/kaj/2015/v1.i26.6140

Alalaq, A. (2013). Mr. Hasan Modarres and his political role in Iran 1870-1937 AD. *Kufa Journal of Arts*, *1*(17), 319–360.
https://doi.org/10.36317/kaj/2013/v1.i17.6445

Scientific researcher Google Scholar,

https://2u.pw/ePsji8S

Mustafa Al-Qayed, What is Google Scholar, Scientific Researcher,
https://www.new-educ.com/cest-quoi-google-scholar

Rasha Abdel Salam, 10 skills to learn in Google Scholar,
https://2u.pw/c8NW1XW